\documentclass{aa}  
\usepackage{graphicx}
\usepackage{xcolor}
\usepackage{listings}
\usepackage{enumitem}
\usepackage[export]{adjustbox}
\usepackage{amsmath}
\usepackage{tikz}
\usetikzlibrary{shapes.geometric, arrows.meta, positioning}

\usepackage[varg]{txfonts}                
\usepackage[utf8]{inputenc}               
\usepackage[T1]{fontenc}
\usepackage{natbib}
\usepackage{ulem}

\usepackage[colorlinks=true,
            linkcolor=blue,
            citecolor=blue,
            urlcolor=blue,
            pdfborder={0 0 0}
           ]{hyperref}

\bibpunct{(}{)}{;}{a}{}{,}
\newcommand{\orcid}[1]{}

\begin{document} 

\title{Closed-form approximations of fundamental quantities of Lemaitre-Tolman-Bondi cosmologies from Symbolic Regression}
   
\subtitle{I. Results on the Garcia-Bellido-Haugbølle parameterization}

\author{A. Carvalho\orcid{0000-0002-9301-262X}\thanks{\email{ascarvalho@fc.ul.pt}}\inst{1,2,3}, A. Krone-Martins\orcid{0000-0002-2308-6623}\inst{3,1}, A.~Da Silva\orcid{0000-0002-6385-1609}\inst{1,2},  J.P.~Mimoso\orcid{0000-0002-9758-3366}\inst{1,2},C. B\oe hm\inst{4,5}}

\institute{$^{1}$ Instituto de Astrof\'isica e Ci\^encias do Espa\c{c}o, Faculdade de Ci\^encias, Universidade de Lisboa, Campo Grande, PT-1749-016 Lisboa, Portugal\\
$^{2}$ Departamento de F\'isica, Faculdade de Ci\^encias, Universidade de Lisboa, Campo Grande, PT1749-016 Lisboa, Portugal\\
$^{3}$ Donald Bren School of Information and Computer Sciences, University of California, Irvine, Irvine CA 92697, USA\\
$^{4}$ Sydney Institute for Astronomy, School of Physics, The University of Sydney, NSW 2006, Australia\\
$^{5}$ University of Edinburgh, Higgs Centre for Theoretical Physics, James Clerk Maxwell Building, Edinburgh EH9 3FD, UK}

\date{Received Month Day, Year; accepted Month Day, Year}



\abstract{
We introduce a novel set of analytic approximations for five fundamental functions in spherically symmetric, inhomogeneous Lemaitre–Tolman–Bondi (LTB) cosmologies, derived via Symbolic Regression (SR). 
Focusing on the constrained Garcia–Bellido–Haugbølle (GBH) parameterization, we sampled the four‐dimensional LTB parameter space using the {\tt bubble} LTB numerical code, then applied SR to reconstruct closed‐form expressions for the radial and transverse scale factors \(A_{\parallel}(r,t)\), \(A_{\perp}(r,t)\), the corresponding Hubble functions \(H_{\parallel}(r,t)\), \(H_{\perp}(r,t)\), and the angular diameter distance \(D_{\mathrm{A}}(z)\). 
Our best‐fit formulas reproduce the numerical data with remarkable precision: the relative mean error across all quantities remains below 0.3\%, except for the radial Hubble function, where it reaches 1.4\%. 
These compact expressions facilitate rapid evaluation of LTB predictions—enabling fast parameter scans, likelihood analyses, and model comparisons—without recourse to time‐consuming integrations. We provide explicit coefficients and discuss the domain of validity, demonstrating that SR-driven 
approximations can serve as robust surrogates for exact LTB solutions in both theoretical
investigations and observational confrontations.
}

\keywords{Symbolic Regression -- Cosmology: theory -- Cosmology: large-scale structure of Universe -- Methods: analytical -- Distance scale -- Inhomogeneous cosmologies}

\titlerunning{Closed-form approximations of fundamental quantities of LTB cosmologies from Symbolic Regression}
\authorrunning{A. Carvalho, et al.}

\maketitle
\nolinenumbers
%

\section{Introduction \label{sec:introduction}}

A central task in cosmology is the fast and accurate prediction of key observables, such as distances, growth functions, and power spectra, across wide regions of parameter spaces. These predictions are essential for confronting theoretical models with increasingly precise data. 
However, obtaining them often requires solving systems of coupled, nonlinear differential equations derived from general relativity, fluid dynamics, and high-energy physics.
Such calculations, while indispensable, are computationally demanding, particularly in applications that require repeated evaluations, including Bayesian inference, Markov Chain Monte Carlo (MCMC) analyses, or real-time model testing.

A way to address these challenges is to rely on analytic fitting functions and approximate formulas. Classic examples include the Carroll–Press–Turner approximation for the linear growth factor \citep{Carroll1992}, the BBKS transfer function \citep{BBKS1986}, the nonlinear matter power-spectrum fits \citep{Takahashi:2012em}, and numerous formulas for luminosity and angular-diameter distances in Friedmann–Lemaître–Robertson–Walker (FLRW) cosmologies \citep[e.g.,][]{Pen1999,Wright2006}. These expressions offer rapid and reasonably accurate surrogates for complex integrals and dynamical calculations, enabling efficient parameter exploration and model comparison.

More recently, machine-learning (ML) techniques have emerged as powerful tools for constructing fast model regressors and emulators trained on simulation data. A diverse set of approaches has been explored in this context. Gaussian Process Regression (GPR) provides smooth interpolation with built-in uncertainty quantification and has been widely applied to the nonlinear matter power spectrum, halo properties, and other large-scale-structure observables \citep[e.g.,][]{Heitmann2009,Lawrence2017,EuclidEmulator2021,LucieSmith2018}. Neural networks, including feedforward and deep learning architectures, are particularly effective for modeling highly nonlinear mappings and high-dimensional outputs, such as image-based simulations and weak-lensing observables \citep{camels2021}. 
N-Body simulation volumes have also been emulated using supervised machine learning \citep[see, e.g.,][]{conceicao_2024} and references therein. 
These methods now play a central role in precision large-scale-structure analyses by enabling fast and accurate interpolation across high-dimensional cosmological parameter spaces.

Despite their success, many ML emulators act as black-box predictors: they provide accurate mappings between cosmological parameters and observables, but typically do not yield compact analytical representations. In settings where interpretability, physical insight, or the construction of human-readable surrogates is desirable, this can be a limitation.

Symbolic Regression (SR), in contrast, searches directly over functional forms to identify closed-form analytical expressions that fit the data. Its strengths lie in interpretability, compactness, and the ability to discover human-readable formulas that can be inspected, simplified, and, in favorable cases, connected to underlying physical mechanisms \citep{doi:10.1126/science.1165893,Bongard2007,Gaucel2014,Sun2019,UdrescuTegmark2020,2020arXiv200610782U,2020arXiv200611287C}. 

Early adoptions of symbolic regression in astronomy in the machine-assisted discovery of empirical relationships in large sky-survey data sets \citep{2013MNRAS.431.2371G} and in the first machine-derived closed-form analytical expression for galaxy photometric redshifts \citep{2014MNRAS.443L..34K}  were revealing and demonstrated its potential. 
Further research with SR has therefore acquired significant traction among the scientific community. For instance, \citet{camels2021} search for an expression for the star formation rate density dependent on redshift ($z$), $\sigma_8$, $\omega_m$, and stellar feedback of simulations, combining N--body and hydrodynamic simulations. \citet{2020arXiv200611287C} combine Graph Neural Networks with Genetic Symbolic regression in dark matter simulations to extract the ``concentration of dark matter from the mass distribution of nearby cosmic structures''. \citet{2023ApJ...956L..22D} explore the $M_\bullet$–$\phi$–$v_\text{max}$ scaling relation plane for Spiral Galaxies with PySR. Also, some recent SR methods are being developed, tailoring specific needs from science, as the core concept that in most research, the datasets come from a combination of experiments, each one using different experimental set-ups \citep[e.g.,][]{10.1145/3638529.3654087}.

These developments have also motivated the application of SR to cosmological inference 
\citep[e.g.,][]{Bartlett2022_Hz_DL,Carvalho2023_sigma8,Castelao2025_cosmogen} and to reconstruction problems 
\citep[e.g.,][]{Bernal2021_nuLIM,2020arXiv200611287C,Aizpuru2021,Orjuela-Quintana2022_Tk,Delgado2022_corrfunc,Shao2022_halos,Wong2022_GW,Shao2023_OmegaM,Wadekar2023_SZ,Wadekar2023_SZ-YM,Wedekar2020_21cm}, where the goal is to obtain accurate yet interpretable closed-form approximations to key functions and observables. Related efforts include the reconstruction of cosmological functions and their use in consistency tests of the standard FLRW cosmology 
\citep[e.g.,][]{Bogdanos2009,Nesseris2010,Nesseris2012,Nesseris2013,Sapone2014,Martinelli2020_euclid,Arjona2020a,Arjona2020b,Arjona2020c,Arjona2021,Arjona2021_GW_lens,Arjona2021_Copernican,Arjona2022,Bartlett2022_Hz_DL,Nesseris2022_euclid,Arjona2024,Ocampo2025_euclid}.

This strategy is particularly relevant for cosmological models whose theoretical predictions are numerically demanding, including inhomogeneous cosmologies that relax the assumption of large-scale homogeneity.

In spite of the observational success of FLRW models, inhomogeneous cosmologies such as the Lemaître–Tolman–Bondi (LTB) class remain highly relevant in modern cosmology. They not only provide a framework to test the robustness of the FLRW assumption of large-scale homogeneity, by accommodating radial inhomogeneities in a fully relativistic setting, but also provide a practical way to investigate the impact of cosmic structures (voids, over-densities, and large-scale gradients) on light propagation, distance measures, and parameter inference, which also help to assess potential back-reaction effects \citep{Buchert:1999pq,Bolejko:2016qku, Buchert:2011sx,Sussman:2011na}. 
Moreover, LTB models have long been studied as alternatives to dark energy in an attempt to describe the late-time acceleration of the universe, or to investigate non-standard explanations of late-time cosmological observations 
\textcolor{purple}{ \citep{Tomita:2000jj,Larena:2008be,Rasanen:2006kp,Enqvist:2007vb,Marra:2011ct}}. 
They have also been used as toy models to study relativistic effects that are hard to isolate in perturbative methods. These include redshift drift, relativistic corrections to structure growth, and light-cone observables. For these reasons, LTB cosmologies remain an active and relevant area of research \citep{Buchert:1999pq,Clarkson:2009sc,Sussman:2010zp,Herrera:2010xr,Herrera:2011kd,Bull:2011wi,Buchert:2015wwr,Buchert:2019mvq,Euclid:2022ucc}, 
especially at a time when observations are becoming increasingly precise and require careful testing of our assumptions about the Universe.

From a practical point of view, one limitation of working with LTB cosmologies is their high computational cost. Evaluating observables for a given set of model parameters is numerically expensive. Physical quantities in LTB spacetimes are usually obtained by solving several coupled and nested ordinary differential equations for the transverse and radial scale factors, $a_1(r,t)$ and $a_2(r,t)$. This must be done together with model-defining radial functions, such as the curvature profile, mass function, and possible variations in the bang-time. In addition, computing observables along the past light cone requires the numerical integration of non-trivial geodesic and light-cone equations. These calculations are highly sensitive to changes in model parameters and often need dense sampling to achieve sufficient accuracy. In practice, modern LTB solvers, such as the \texttt{bubble} code \citep{phillBullWiki} used in this work or LLTB codes like VD2020 \citep{valkenburg_vd2020,2012GReGr..44.2449V}, typically take from tens of seconds to a minute to evaluate a single point in parameter space on a standard laptop. This computational cost makes full Bayesian inference impractical. Markov Chain Monte Carlo or nested-sampling methods would require tens to hundreds of thousands of solver evaluations, which is not feasible in high-dimensional parameter spaces. For this reason, there is strong motivation to develop fast, accurate, and interpretable surrogate models. Examples include analytical approximations based on symbolic regression, which can replace repeated full numerical integrations in cosmological parameter inference.

While ML-based emulators are now common in standard FLRW cosmology, their application to inhomogeneous models, such as the LTB class, remains largely unexplored. LTB models \citep{Lemaitre1933,Tolman1934,Bondi1947} describe spherically symmetric, dust-dominated universes without assuming homogeneity, and have been widely used to study cosmic voids \citep{Tomita:2000jj,2008JCAP...04..003G,Celerier:2009sv,2008PhRvL.101m1302C,Camarena:2022iae},
alternatives to dark energy \citep{Enqvist:2006cg,Buchert:2007ik,Krasinski:2010pfm,Bolejko:2008yj}, 
and the effects of inhomogeneities and backreaction \citep{Ishibashi:2005sj,Clarkson:2011zq,Valkenburg:2012td,Redlich:2014gga,Green:2014aga,Buchert:2015wwr,Larena:2008be,Bester:2015gla,Fleury:2016fda,Camarena:2021mjr}. 
Their numerical implementation, however, is considerably more involved than in FLRW cosmology, requiring the solution of several nested differential equations for each parameter set. This complexity has limited their use in large-scale cosmological inference, where fast evaluations are crucial.

In this work, we present what is, to the best of our knowledge, the first application of Symbolic Regression to LTB cosmologies. Focusing on the constrained García-Bellido–Haugbølle (GBH) subclass of LTB models \citep{2008JCAP...04..003G}, we generate a large dataset using the \texttt{bubble} LTB code \citep{phillBullWiki} and train an SR algorithm to obtain compact analytical expressions for five fundamental quantities: the radial and transverse scale factors, their associated Hubble rates, and the angular-diameter distance. These SR-derived formulas provide a computationally efficient and interpretable alternative to full numerical integration, typically achieving relative mean errors below 0.3\% for most quantities, and ~338\% improvement in running time compared to the \texttt{bubble} code.

The structure of this paper is as follows. In Section~\ref{sec:theory}, we briefly review the LTB formalism and the constrained GBH parameterization. Section~\ref{sec:SRmethod} introduces the Symbolic Regression method and our dataset generation process. Section~\ref{sec:results} presents the resulting analytical formulas and evaluates their accuracy. Finally, Section~\ref{sec:conclusions} summarizes our findings and discusses possible applications and extensions.

\section{The Cosmological Model \label{sec:theory}}
\subsection{Lemaitre Tolman Bondi cosmological model}
\label{sec:LTB}

The LTB model is derived from a spherically symmetric solution to Einstein's equations, describing an isotropic yet inhomogeneous universe. This model excludes the contribution of dark energy, with matter density and curvature density as its primary components. Unlike the FLRW metric, where the scale factor and the Hubble parameter are time-dependent only, the LTB model introduces radial dependencies for these factors. In fact, the LTB metric, 
\begin{equation}
    ds^2 = dt^2 - \frac{a^2_2(t,r)}{1-k(r)r^2} dr^2 - a^2_1(t,r)r^2 d\Omega^2 .
\end{equation}
introduces two metric functions, $a_1(t,r)$ and $a_2(t,r)$ related by\footnote{\label{derivative_notation} In this work, a dot denotes the partial time ($t$) derivative, while the apostrophe symbolizes the partial space ($r$) derivative.}
\begin{equation}
    a_2 = (a_1 r)',
    \label{eq:scale_factor_ltb}
\end{equation}
where $a_1$ satisfies a Friedmann-like equation:
\begin{equation}
    \left( \frac{\dot a_1}{a_1} \right)^2 = \frac{8 \pi G}{3} \frac{m(r)}{a_1^3} - \frac{k(r)}{a_1^2}\; ,
    \label{eq:friedman}
\end{equation}
where $m(r)$ is related to the so-called Misner-Sharpe mass $m_{ms}=4\pi\,\int\,\rho(r)\,r^2\,{\rm d}r$ with $\rho(r)$  being the energy-density at some fiducial initial time.

These allow to define transverse and radial Hubble rates as:
\begin{eqnarray}
 H_T \equiv \frac{\dot{a_1}}{a_1} \label{eq:HT} \\
 H_R \equiv \frac{\dot{a_2}}{a_2} \label{eq: HR}.
  \end{eqnarray}
For an observer at the centre of the LTB non-homogeneity,  the following differential equations describe the relationships between radial ($ds^2 = d \Omega^2 = 0$) geodesic properties, time, and redshift:
\begin{eqnarray}
\frac{dt}{d \log (1+z)} = - \frac{a_2}{\dot a_2} = - \frac{1}{H_R},  \label{HT} \\
\frac{dr}{d \log (1+z)} = \frac{\sqrt{1- k(r)\,r^2}}{\dot a_2} \label{HR}.
\end{eqnarray}
Using these equations, one can calculate the angular diameter distance, $D_A$, and luminosity distance $D_L$ as:
\begin{eqnarray}
&&D_A (z)=r(z) \, a_1 \left(r(z), t(z) \right), \label{eq:da_LTB} \\
&&D_L (z)=(1+z)^2 \, D_A(z), \label{eq:dL_LTB} 
\end{eqnarray}
where $D_A$ is the angular distance and $D_L$ the luminosity distance.

When describing an inhomogeneous universe, it is essential to specify the LTB inhomogeneous profile. In this work, we adopt the {\it constrained} GBH model \citep{2008JCAP...04..003G}, described in the next section. 

\subsubsection{Garcia-Bellido and Haugbølle model}
\label{subsec:GBH}
The GBH model \citep{2008JCAP...04..003G} describes a central inhomogeneity characterized by the present matter density $\Omega_M(r)$ and expansion rate $H_0(r) $ profiles. The $\Omega_M(r)$ is given by:
\begin{equation}
    \Omega_M(r) = \Omega_{\text{out}} + \Big( \Omega_{\text{in}} - \Omega_{\text{out}} \Big) \Bigg( \frac{1 - \tanh{[(r-r_0)/2\Delta r]}}{1 + \tanh{[r_0/2\Delta r]}}\Bigg).
    \label{eq:GBH_omega_profile}
\end{equation}

In this work, we adopt a more constrained and commonly used version of the LTB model, known as the constrained LTB model, in which the time of the Big Bang, $t_{BB}$, is constant 
for all observers, regardless of their spatial position. This constraint is achieved by imposing the following condition:
\begin{equation}
H_0(r) = H_0\left[{1\over \Omega_K(r)} -
{\Omega_M(r)\over\sqrt{\Omega_K^3(r)}}\ {\rm sinh}^{-1}
\sqrt{\Omega_K(r)\over\Omega_M(r)}\right]\,,
\end{equation} 
where $\Omega_K(r)=1-\Omega_M (r)$. Here $H_0=H_0(r \to \infty )=100 \, h $ km s$^{-1}$ Mpc$^{-1}$ and $h$ is the dimensionless Hubble parameter. 
This constrained version of the model has reduced flexibility compared to the GBH model, as it includes only a single arbitrary function, $\Omega_M(r)$, and one fewer free parameter. Additionally, we adopt the popular assumption of fixing $\Omega_{\text{out}} = 1$, which simplifies the model further. With this choice, the model is parameterized by only four parameters: $h$, $\Omega_{in}$, $r_0$ and $\Delta_r$.

\section{Methodology \label{sec:SRmethod}}

In this section, 
we briefly describe \texttt{bubble}, an LTB cosmology calculator that was employed for the mocks generated (as per in Sec.~\ref{sec:ltb_data}). Afterward, we introduce Symbolic Regression, a machine-learning method designed to find approximated analytic expressions of relations between features in a data set. We also describe TuringBot, the SR engine we adopted to search closed-form analytical expressions that relate the LTB cosmological functions to the corresponding cosmological parameters, and we finally describe how we adapted it to explore the large parameter space.

\subsection{\texttt{bubble}}
\label{sec:bubble}

\texttt{bubble} \footnote{Source code can be accessed at: \texttt{https://gitlab.com/cosmobubble/bubble}.} is a numerical code for LTB models cosmology, written in C++. It calculates geodesic and background properties such as the scale factor, as well as observables like the kinetic Sunyaev-Zeldovich effect, the luminosity distance, and the redshift, for a set of theoretical parameters of the chosen LTB model. The code incorporates two void profile models: the CFL model \citep{2008PhRvL.101m1302C} and the previously mentioned GBH model (referenced in Sec. \ref{subsec:GBH}), and it also allows users to implement their own void model. Additionally, an FLRW metric is included in this cosmological package. 
Given a set of parameters for the void model, \texttt{bubble} computes the geodesic quantities and observables across a range of radial coordinates \(r\) and outputs these quantities in \(r\)-dependent tables. Therefore, extrapolations are necessary if the desired value falls between two tabulated values. 
The cosmological properties we are primarily interested in, such as \(H_r\), \(H_t\), \(a_1\), and \(a_2\) as functions of \(z\), are obtained through the \texttt{output\_fnz} function, declared within the \texttt{model\_output.cpp} file.

\subsection{Symbolic Regression}
\label{sec:SR}
One major obstacle to using machine learning methods, especially neural networks, is that the resulting models are extremely hard to interpret and thus are often treated as ``black boxes.'' This is a serious limitation in a physics-driven field like astrophysics, where we ideally want to understand and test the mapping from observables to the target quantities. In contrast, symbolic regression (SR) can overcome this drawback, as the resulting expressions are explicit and can be directly interpreted. In fact, the SR method AI Feynman \citep{2020arXiv200610782U}, was able to re-discover over 100 equations from ``Feynman lectures on physics''\citep{1963flp..book.....F, 1964flp..book.....F, 1965flp..book.....F}. 

SR is an ML method that aims to generate mathematical symbolic expressions that can describe a given data set. The method itself is historically based on Genetic Programming \cite{koza_genetic_1992, koza_genetic_1994}, but current methods implement a large variety of optimization algorithms \citep[e.g.,][]{2021arXiv210714351L}, classic neural networks \cite[e.g.,][]{landajuela_unified_2022}, and more recently, Generative AI heuristics based on transformers \cite[e.g.,][]{kamienny_end--end_2022, 10.5555/3666122.3668112}. 

Traditionally, SR will search for the combination of mathematical operations that minimizes an error while promoting the simple equations it found. A common representation of an SR method is a tree, where each node is a mathematical expression, a numeric constant, or a data feature/input. The method aims to explore every combination possible of nodes. Currently, there is a broad offer of SR algorithms. Some of the best-performing ones, according to their accuracy when compared to the ground-truth process, ranked by the benchmarking tool SRBench \citep{2021arXiv210714351L} are: MRGP \citep{10.1145/2576768.2598291}, Operon \citep{10.1145/3377929.3398099}, the already mentioned AI Feynman, SBP-GP \citep{10.1145/3321707.3321758} and PySR \citep{2023arXiv230501582C}. A well-known closed-source software is TuringBot (TB), inspired by 
the pioneering -- but now discontinued-- Eureqa \citep{doi:10.1126/science.1165893} method. In this work, we will use the TuringBot, which we further describe in Sec.~\ref{sec:TB}.

\subsubsection{The Turing Bot engine}
\label{sec:TB}
Here, we use the Turing Bot method (TB; \citealt{TuringBotDoc}), which is an SR method implemented in C++ and based on Simulated Annealing. Early results obtained using TB \citep{2020arXiv201011328A} indicated that for physics-inspired problems, it performed better than Eureqa, which was a popular and widely successful SR system \citep{doi:10.1126/science.1165893}. 

Due to being a multithreaded C++ engine, TB has been highly efficient compared to many of the published SR  implementations, which allows it to explore complex datasets. The method has been previously used in multiple contexts in physics and engineering \citep[e.g.,][]{2021arXiv211112210L}, including in astrophysics \citep{2022ApJ...940...30B}. A practical limitation of TB, though, is that it is a closed-source system. This, in part, restricts what can be tailored for any specific adoption of TB, such as the underlying error metric, which can only chosen from those already implemented. On the other hand, it profits from a highly stable implementation, and this simplifies deploying the TB engine in highly distributed environments, thus enabling systematic exploration of large parameter spaces. This robustness and stability are the main reasons why we adopt TB in this work. Moreover, the interaction with the engine can be performed directly via configuration and output files through the command line, which makes it straightforward to script and automate large batches of runs.

The TB configuration file allows the selection of the error metric used in the minimization step, as well as specifying an explicit formula for the input features to be used by the method. Within the same file, the user can set the maximum expression complexity for TB to explore, the train–test split ratio for cross-validation, the number of best equations $n$ to be written to the output, and the set of allowed functions that may appear in the SR expression trees, among other options. TB offers a graphical user interface, which we do not use here, but it can also be executed in a fully non-interactive mode from the command line by passing the configuration file as input.

The output of TB is a text file that is created at the start of the run and that is continuously updated as the engine is running. This file contains the current iteration's best $n$ equations, with $n$ chosen by the user, together with metrics showing the expression algebraic complexity and the value of the selected error metric for each candidate symbolic regression solution. 

\subsubsection{Exploring the parameter space}
Due to its multithreaded implementation, TB can be efficient when executed on a multicore environment. Nevertheless, as it is, the TB implementation does not scale to multinode environments. Moreover, depending on characteristics of the data to be explored by the method, especially its volume and the number of features (possibly leading to memory transfer inefficiencies between the threads), and the complexity of the equations to be searched, instead of running a single TB process over all cores of a given computer it can be more efficient to run a larger number of TB processes each one using a smaller number of cores and starting with different seeds, thus starting from different parts in the parameter space. 

To enable the use of TB and other SR engines in scenarios involving large data sets and/or multi-node clusters and supercomputers, and to possibly lead to a wider exploration of the parameter space, we have developed the Orchestrator tool (see Appendix~\ref{appendix:code} for more details). Orchestrator is responsible for launching, monitoring, and coordinating multiple parallel symbolic regression processes. It currently supports TB and, in more recent versions, also PySR. 

In this work, we used Orchestrator with the TB engine in the Expanse system, a cluster located in the San Diego Supercomputer Center (SDSC). For our experiments, we used up to 128 CPU cores. But the Expanse system has 728 compute nodes based on AMD EPYC 7742 CPUs with 64 cores each, and 52 GPU nodes with a total of 208 V100 GPUs. It provides 247~TB of total system memory and 824~TB of flash storage. To obtain the reconstructed LTB cosmology equations from SR presented in the next section, we used a total of $39\,528$ CPU hours, excluding the experiments that led to the development of Orchestrator.
\vspace{5mm}

\section{Results \label{sec:results}}

In this section, we will present the analytical expressions achieved by the application of Symbolic Regression for the LTB-GBH model. We start by describing the data creation process, followed by the obtained approximated equations for $H_R$, the radial Hubble rate, $H_T$, the transverse Hubble rate, $a_1$ and $a_2$ components of the scale factor, $D_A$, the angular diameter distance, and $dV/dz$, the comoving volume element.

\subsection{Building the Training Data}
\label{sec:ltb_data}

\begin{figure*}
    \centering
    \includegraphics[width=1.8\columnwidth]{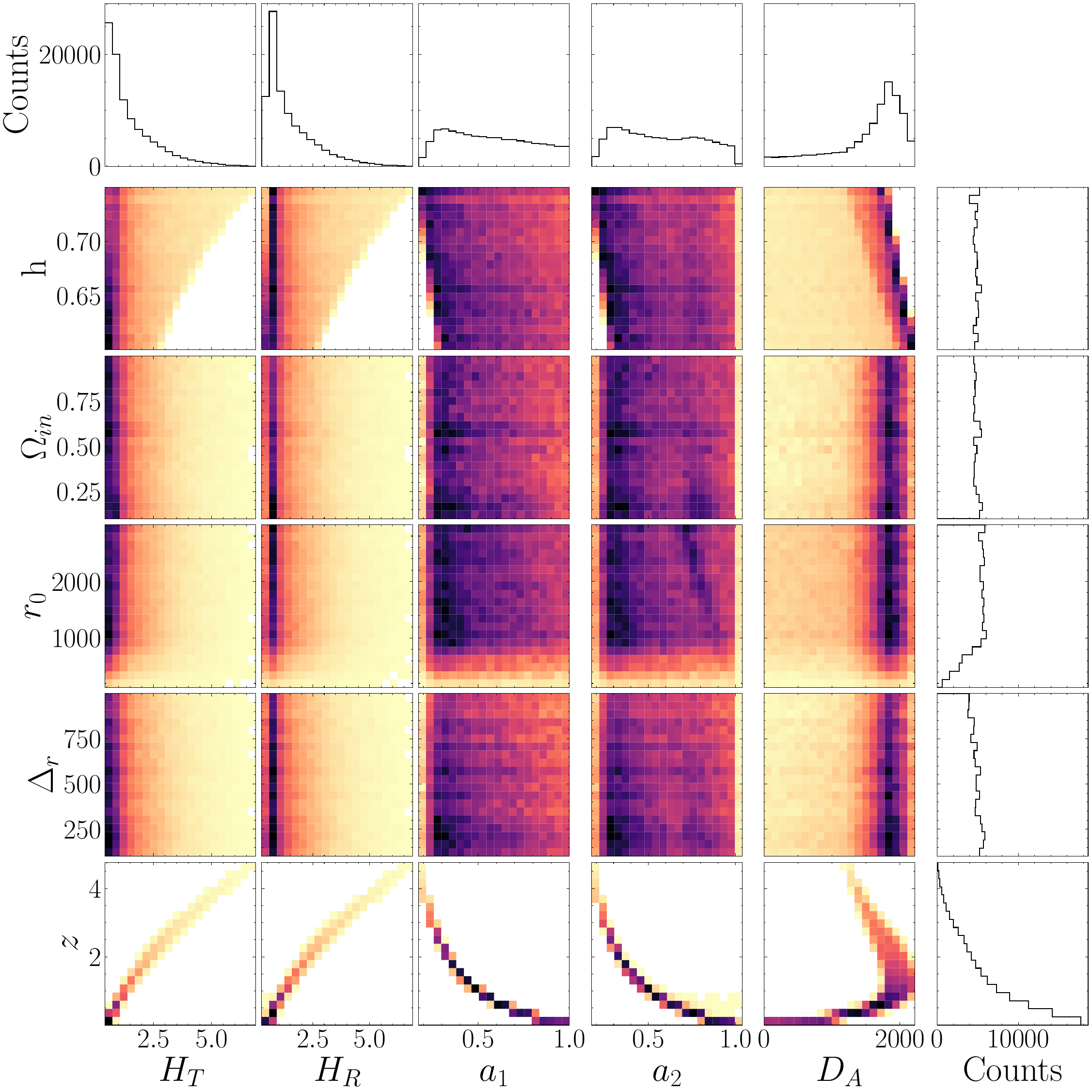}
    \caption{2D Distribution of the cosmological parameters generated from \texttt{bubble} as described in Sec.~\ref{sec:ltb_data}, showing the relation between the input parameters (rows) and the target output quantities (columns). In here, the most darker color corresponds to regions that are denser in points, whereas the lighter represents regions with sparser points, where the white color indicates lack of population. Additionally, the histograms for all quantities here represented are shown in the first row for the target functions and the last column for the input cosmological parameters of our Symbolic Regression methodology.}
    \label{fig:all_scatter}
\end{figure*}

We are interested in deriving analytical expressions for the parameters of a inhomogeneous cosmological model. In particular, we are considering the Lemaitre-Tolman-Bondi (LTB) model combined with a GBH void profile (Sec. \ref{subsec:GBH}). This model can be characterized mainly by 4 parameters:
$h$ the Hubble parameter, $\Omega_\text{in}$ the matter density inside the inhomogeneity, $r_0$ the size of the inhomogeneity region and $\Delta_r$ the transition to uniformity. Thus, we have randomly generated 5000 sets (hereafter models) described by these 4 cosmological parameters inside the following intervals:
\begin{align}
\begin{aligned}
    h = [0.60, 0.75]  && \Delta_r = [0.1,1] \, \text{Gpc} && \\
    \Omega_\text{in}= [0.1,1] && r_0= [0.1,3] \, \text{Gpc}, &&
\end{aligned}
\end{align}
where we conditionally sampled $r_0$ as such that $r_0 > \Delta_r$. We used \texttt{bubble} (Sec.~\ref{sec:bubble}) to compute the numerical values of the functions we are interested in learning an approximation analytical function: $H_R$, $H_T$, $a_1$, $a_2$, and $D_A$. For each parameter model, the code produces a table of cosmological background quantities as a function of redshift. For each table, we randomly select 20 rows to be able to have a redshift dependency on our target cosmological functions, as it is an important observable in cosmology. In Fig.~\ref{fig:all_scatter} we show both the distribution of the generated features (right, rotated histograms) and the distribution of the response cosmological products (top histograms). In the panels surrounded by the two types of histograms, we display the 2D distribution between the features and the values of the respective theoretical functions. In the ideal scenario, one would generate parameters that are approximately uniformly distributed and can cover the relevant range within the target functions. In our case, we can conclude that our parameter space shows a uniform scatter across the area of interest. This figure also shows us that, as expected, the redshift is highly important in the behavior of the functions to study, as one would conclude from the spatial dependency in their definition, written in Sec.~\ref{sec:LTB}.
The data was split into training and test sets. The training data, consisting of $80\%$ of the 5000 models, will be used as input in the TuringBot software, while the remaining $20\%$ will be used in the test data for the comparison criteria that commands the number of iterations, as described in Appendix~\ref{appendix:code}.

In this study, the configuration file for TB was identical across all explored quantities. The error minimization metric used was the root mean square error (RMSE), so that the large errors are weighted higher in the tool. For the training and testing dataset split, a value of -1 was used for the \texttt{train\_test\_split} parameter, which means no cross-validation was performed as the Orchestrator already considers a train/test split. The use of integer constants was disabled (\texttt{integer\_constants} = 0), so as not to force the constants to be integers. Here, we also deactivated the \texttt{bound\_search\_mode} parameter, meaning the model did not attempt to search for upper or lower bounds of the target parameter, in order to reduce the complexity of the formulas to be found. To control the complexity of the generated formulas, we chose a maximum formula complexity of 60. This limits the number of terms or depth in the equations, reducing overfitting. The \texttt{history\_size} parameter was set to 20, meaning only the top 20 best-performing formulas were stored, ensuring only the most relevant solutions were considered. The F-score beta parameter was set to 1, balancing precision and recall at the same level. We also disabled data normalization transformations. Additionally, the target parameter was not allowed in the lag functions (\texttt{allow\_target\_delay} = 0), as this application is not a time-series forecast. The model was also set not to force the inclusion of all variables (\texttt{force\_all\_variables}=0), ensuring only the most significant variables for each case were used in the formulas. Lastly, the allowed functions included basic arithmetic operations, exponential, logarithms, square roots, and hyperbolic functions. The inclusion of hyperbolic functions was chosen because they can be expressed in terms of exponential and not because a periodic behavior is expected.
These settings were selected to balance model complexity, computational efficiency, and the generalization ability of the generated formulas, ensuring that the results were both accurate and interpretable.

\begin{table}[]
\centering
\begin{tabular}{lcccc}
\hline
\multicolumn{1}{c}{\textbf{\begin{tabular}[c]{@{}c@{}}Color \\ Label\end{tabular}}} &
$\boldsymbol{h}$ &
$\boldsymbol{\Omega_{\text{in}}}$ &
$\boldsymbol{r_0}$ &
$\boldsymbol{\Delta_r}$  \\ \hline \hline
Blue   & 0.65 & 0.18 & 1315.14 Mpc & 652.52 Mpc \\
Green  & 0.60 & 0.40 & 1604.56 Mpc & 586.46 Mpc \\
Orange & 0.70 & 0.62 & 1899.27 Mpc & 651.31 Mpc \\
Red    & 0.71 & 0.33 & 2212.78 Mpc & 588.09 Mpc \\
Purple & 0.66 & 0.33 & 2883.71 Mpc & 665.01 Mpc \\ \hline \hline
\end{tabular}
\caption{Cosmological model parameterization considered in the comparison plots of Fig.~\ref{fig:SR_curves-1} and Fig.~\ref{fig:SR_curves-2}.}
\label{tab:LTB_models}
\end{table}

\subsection{Analytical Approximations of LTB functions}

Below, we introduce the cosmological equations derived using the Orchestrator. The cosmological quantities considered for constructing these functions include $h$, $\Omega_{\text{in}}$, $r_0$, $\Delta_r$ and $z$. It is important to note that some functions may not include all parameters. This omission occurs when TuringBot does not identify a sufficiently strong relationship between the excluded parameter and the target function.

\begin{figure*}
    \centering
    \includegraphics[width=0.73\columnwidth]{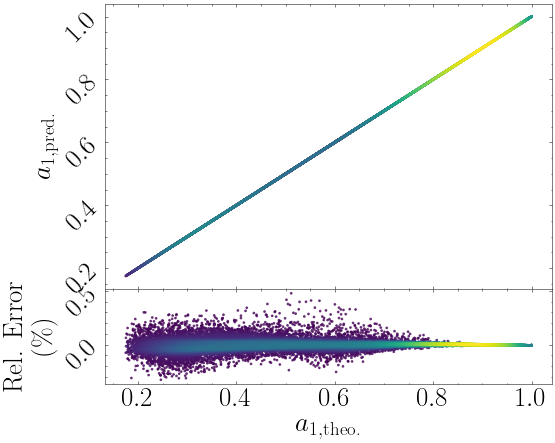} \hspace{1cm}
    \includegraphics[width=0.7\columnwidth]{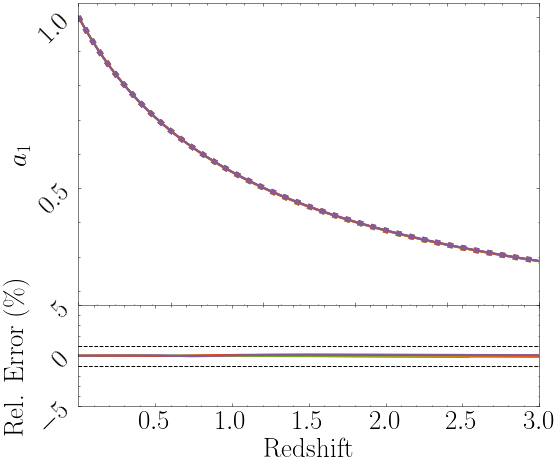}
    \includegraphics[width=0.73\columnwidth]{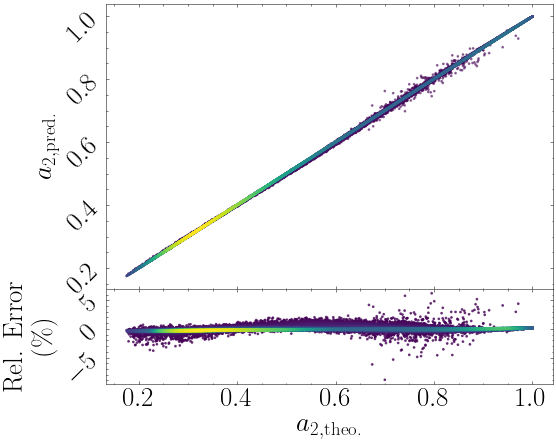}\hspace{1cm}
    \includegraphics[width=0.7\columnwidth]{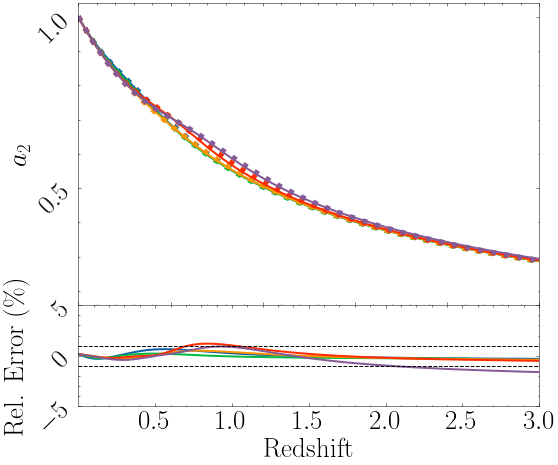}
    \caption{Representation of the SR found expressions of the transverse scale factor $a_1$ and the radial scale factor $a_2$. Left panel: Comparison between the theoretical values obtained via \texttt{bubble} software and the estimated values from the Orchestrator, where a yellow color represents the most dense regions. In the bottom panels of each left panel, we showcase the relative error between these two values. Right panel: Redshift distribution of each parameter for the models labeled in Tab~\ref{tab:LTB_models}. The dashed line represents the theoretical distribution from \texttt{bubble} and the solid line represents the redshift distribution obtained from the estimated equation.}
    \label{fig:SR_curves-1}
\end{figure*}

\begin{figure*}
    \centering
    \includegraphics[width=0.72\columnwidth]{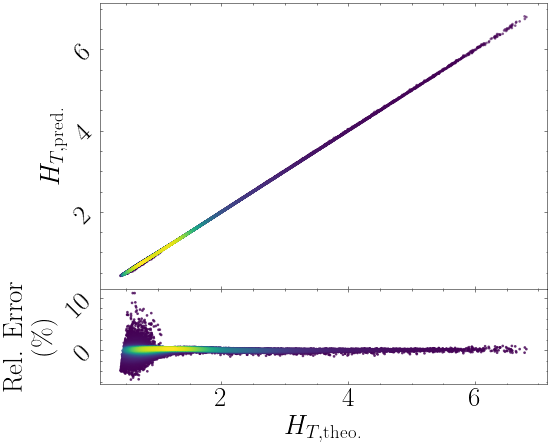} \hspace{1cm}
    \includegraphics[width=0.7\columnwidth]{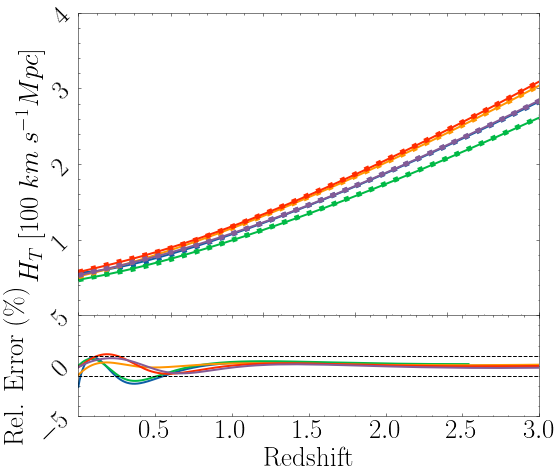}
    \includegraphics[width=0.75\columnwidth]{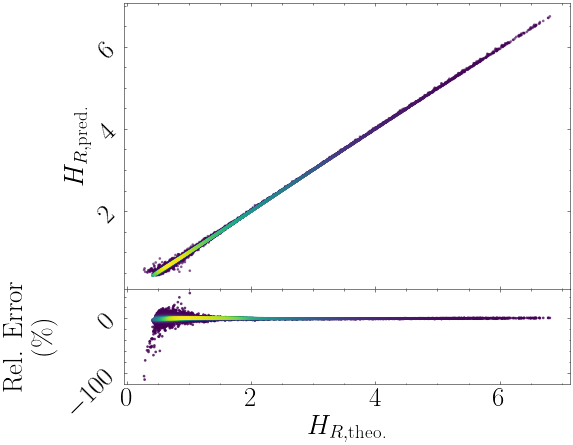}\hspace{1cm}
    \includegraphics[width=0.7\columnwidth]{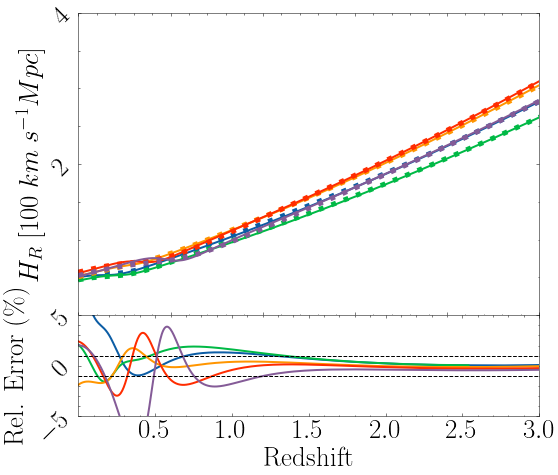}
    \includegraphics[width=0.69\columnwidth]{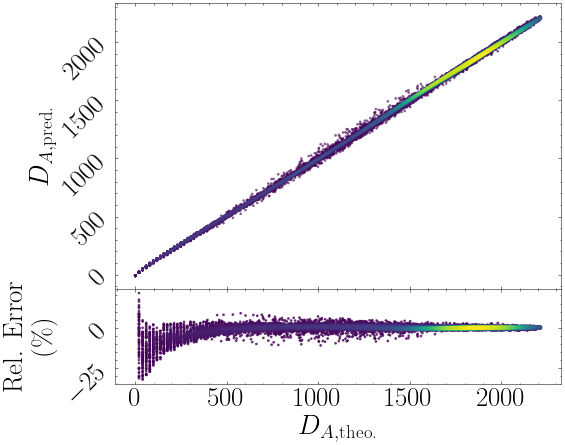}\hspace{1cm}
    \includegraphics[width=0.7\columnwidth]{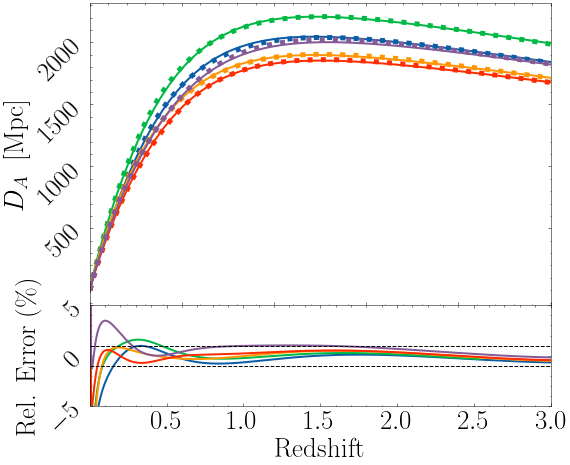}
    \caption{Representation of the SR found expressions of the transverse Hubble rate $H_T$, the radial Hubble rate $H_R$, and the angular diameter distance $D_A$. Left panel: Comparison between the theoretical values obtained via \texttt{bubble} software and the estimated values from the Orchestrator, where a yellow color represents the most dense regions. In the bottom space of each left panel, we showcase the relative error between these two values. Right panel: Redshift distribution of each parameter for the models labeled in Tab.~\ref{tab:LTB_models}. The dashed line represents the theoretical distribution from \texttt{bubble} and the solid line represents the redshift distribution obtained from the estimated equation.}
    \label{fig:SR_curves-2}
\end{figure*}

In Fig.~\ref{fig:SR_curves-1} and Fig.~\ref{fig:SR_curves-2} the left-hand-side plots display the true values obtained from \texttt{bubble} code alongside the corresponding estimated values for the cosmological parameters related to the models. The bottom panels in these plots illustrate the relative difference between the estimated curves and the theoretical curves. In the right-hand-side plots, we present the redshift distribution of the models listed in Tab.~\ref{tab:LTB_models}, where the 
show the small symbols
represent the true curve, and the solid lines indicate the curve derived from the equations obtained through the SR process. The bottom panels again show the relative error between the theoretical and the estimated curves. Additionally, in Tab.\ref{tab:tab_errors}, we provide the average and median values of the relative error $(\text{Pred.} - \text{Theo.})/\text{Theo.} \times 100\%$ for the test set concerning each cosmological function and the percentage of points above a certain relative error threshold.

The constants $C_i$ that appear in all reconstructed functions of this section are presented in Tab.~\ref{tab:SR_constants}. Moreover, in the functions of this section, the cosmological parameters $r_0$ and $\Delta_r$ are in Mpc.

\begin{table*}[h!]
\centering
\begin{tabular}{lccccccc|cc}
\hline
\textbf{Threshold} & \textbf{$\pm$0.1\%} & \textbf{$\pm$0.2\%} & \textbf{$\pm$0.5\%} & \textbf{$\pm$1.0\%} & \textbf{$\pm$2.0\%} & \textbf{$\pm$5\%} & \textbf{$\pm$10\%} &  Average Rel. Error & Median Rel. Error \\
\hline\hline
$a_1$  & 0.9294 & 0.9902 & 1.0000 & 1.0000 & 1.0000 & 1.0000 & 1.0000 & $0.0009\%$      & $0.0011\%$        \\ 
$a_2$  & 0.2783 & 0.5267 & 0.8697 & 0.9574 & 0.9945 & 0.9997 & 1.0000  & $0.0477\%$      & $0.0917\%$        \\ 
$H_T$ & 0.1818 & 0.3611 & 0.7396 & 0.8666 & 0.9516 & 0.9979 & 0.9999 & $0.037\%$       & $-0.0566\%$       \\ 
$H_R$  & 0.1177 & 0.2216 & 0.4494 & 0.6305 & 0.7916 & 0.9429 & 0.9871 & $1.410\%$      & $0.6083\%$         \\ 
$D_A$  & 0.1162 & 0.2268 & 0.5517 & 0.7981 & 0.8931 & 0.9542 & 0.9815 & $0.3654\%$      & $0.0105\%$        \\
\hline\hline
\end{tabular}
\caption{Columns 2 to 7 show the fraction of points reconstructed with relative errors less than $\pm0.1\%$, $\pm0.2\%$, $\pm0.5\%$, $\pm1.0\%$, $\pm2.0\%$, $\pm5\%$, and $\pm10\%$ for each symbolic regression function listed in the first column. Columns 8 and 9 report the average and median relative errors, respectively, for each case.}
\label{tab:tab_errors}
\end{table*}

\begin{table*}
\centering
\begin{tabular}{|lccccc|}
\hline
\multicolumn{1}{|c}{}        & $a_1$                                                & $a_2$       & $H_t$        & $H_r$                                                 & $D_A$         \\ \hline
\multicolumn{1}{|l|}{$C_1$}  & 0.80694                                            & 0.58403     & 0.45977     & 0.61237                                           & $0.51710$       \\
\multicolumn{1}{|l|}{$C_2$}  & 0.80687                                             & 0.58501   & 0.24540    & 0.20390                                             & $0.943412$  \\
\multicolumn{1}{|l|}{$C_3$}  & 0.80764                                             & 0.58071   & 1.33346      & 1.33623                                               & $0.279151$      \\
\multicolumn{1}{|l|}{$C_4$}  & $4.23582 \times 10^{-7}$ & $1.44156 \times 10^{-2}$   & $9.99509\times 10^{-2}$    & $3.63985\times 10^{-5}$  & $5.7740\times 10^{-4}$    \\
\multicolumn{1}{|l|}{$C_5$}  & $7.32962 \times 10^{-2} $                                       & $3.84319\times 10^{3}$     & 0.20482      & $2.60339\times 10^{-5}$   & $3.96045$ \\
\multicolumn{1}{|l|}{$C_6$}  & 0.19531                                              & $2.95325\times 10^{-4}$ & $-1.27188\times 10^{-4}$ & $0.90316 $                                           & $1.457922\times 10^{-4}$  \\
\multicolumn{1}{|l|}{$C_7$}  & $2.02548 \times 10^{-4}$                                          & $6.92104\times 10^{2}$     & 4.85370       & 0.02795                                             & -0.943412     \\
\multicolumn{1}{|l|}{$C_8$}  & -                                                    & 75.9261     & $7.00377\times 10^{-4}$  & $0.13744$                                          & 0.90973       \\
\multicolumn{1}{|l|}{$C_9$}  & -                                                    & 0.27963    & 1.03388      & $2.00312\times10^{-4}$                                              & $4.41448\times 10^{-2}$    \\
\multicolumn{1}{|l|}{$C_{10}$} & -                                                    & -           & -            & $2.49206\times10^{-4}$                                             & $1.37548\times 10^{2}$       \\\hline
\end{tabular}
\caption{Constants of the equations found by the Symbolic Regression process for $a_1$ (Eq.~\ref{eq:sr_a1}), $a_2$ (Eq.~\ref{eq:sr_a2}), $H_t$ (Eq.~\ref{eq:sr_ht}), $H_r$ (Eq.~\ref{eq:sr_hr}) and $D_A$ (Eq.~\ref{eq:sr_da}).}
\label{tab:SR_constants}
\end{table*}

\subsubsection{Transverse scale factor}
The LTB scale factors are fundamental quantities of the model, 
as many cosmological and astrophysical functions depending on them. One notable example is the various definitions of distance, such as the angular diameter distance for LTB, which is given in Eq.~(\ref{eq:da_LTB}). We thus begin by exploring an approximated expression for the transverse scale factor $a_1$ of the LTB framework, which characterizes the expansion rate in a perpendicular direction from the line of sight of the observer:

\begin{align}
    a_1 = \frac{C_1}{C_2+z\cdot \left(C_3-\frac{C_4 r_0}{C_5+C_6\frac{ \Omega_\text{in}}{\sqrt{1-\Omega_\text{in}^2}}-C_7 r_0+z}\right)},
    \label{eq:sr_a1}
\end{align}

Recalling how the scale factor is written in a FLRW cosmology, $a(z) = 1/(1+z)$, it is interesting to note that we have obtained an expression for the transverse scale factor that has a similar dependence on the redshift, showing that the method is consistent with the leading order behavior, with the term in the parenthesis acting as a ``correction term''. Moreover, to preserve the validity of the recovered expression in the FLRW limiting case, we would expect that $C1\sim C2\sim C3$, which we indeed see as SR found all these constants to be valued $C_{1,2,3}\sim0.81$. 
Furthermore, the term $\Omega_\text{in}/\sqrt{1 - \Omega^2_\text{in}}$ decodes the inhomogeneity influence of the density profile.
Looking at the constants and what they are associated with, in particular $C_4 = 4.23582 \times 10^{-7}$ and $C_7 = 2.02548 \times 10^{-4}$, it seems TuringBot has found coefficients that normalize its companion feature to balance their contribution to the expression. Do note that we have not performed any data normalization or scaling technique in this work. This highlights the interpretability of the resulting SR expressions.

\subsubsection{Radial scale factor}
With an expression for the transverse scale factor established, we now aim to obtain a corresponding expression for the radial counterpart. The radial scale factor $a_2$ describes the expansion rate along the radial direction. As such, and unlike $a_1$, it varies across the radial coordinate, as seen in Eq.~\ref{eq:scale_factor_ltb}, reflecting the nature of expansion in the LTB framework. The approximated expression found by Symbolic Regression is the following:
\begin{align}
    a_2 = \frac{C_1}{C_2 + z\cdot\left(C_3-\frac{ \left(C_7+C_4r_0\right) \text{sech}\left(\frac{ C_6 h r_0- C_5\tanh (z)}{\Delta_r}\right)}{\left(C_8+\Delta_r\right) \left(C_9+\sinh \left(\frac{\Omega_\text{in}}{\sqrt{1-\Omega_\text{in}^2}}\right)\right)}\right)}.
    \label{eq:sr_a2}
\end{align}

The expression for the radial scale factor is more complex than that for the transverse factor, as expected. Nevertheless, the functional forms of $a_1$ and $a_2$ exhibit a similar structure, reflecting their connection to the same cosmological concept of the scale factor. Moreover, these expressions are closely related to how the scale factor itself is defined by the redshift in FLRW cosmology, where $a(z) = 1/(1+z)$. As in $a_1$, here we also find the term $\Omega_\text{in}/\sqrt{1 - \Omega^2_\text{in}}$, highlighting the inhomogeneity of the density profile.
Again, the values of the constants here reflect a normalization effort by TuringBot. For instance, the constants associated with $r_0$, which is used in Mpc units, $C_4 = 1.44156 \times 10^{-2}$ and $C_6 = 2.95325 \times 10^{-4}$, balance the order magnitude of this distance and its combination with the adimensional Hubble parameter, respectively. Finally, it is interesting to see hyperbolic terms appearing in the $a_2$ expression found by SR, because the GBH matter density contains an hyperbolic term $\tanh[(r-r_0)/2\Delta_r]$, and thus its derivative should introduce $\operatorname{sech}^2$ functions, that then should appear in any analytic approximate expressions for $a_2$, as we are indeed seeing from the $\operatorname{sech}$ term in Eq. \ref{eq:sr_a2}.

\subsubsection{Transverse Hubble rate}
In LTB models, the Hubble rate consists of two components as discussed in Sec.~\ref{sec:LTB}. Hubble rates are fundamental concepts in cosmology and thus contribute to many other cosmological functions.
Therefore, from a computational perspective, it is advantageous to replace the \texttt{bubble} pipeline for Hubble rate retrieval with approximated expressions.
Below, we present the SR expression Orchestrator found for the transverse component $H_T$, the Hubble rate in the perpendicular direction to the observer’s line of sight: 

\begin{equation}
    \begin{aligned}
         \frac{H_T}{100} & =  h \, \Bigg( C_1 +  \left(C_2+z\right){}^{C_3}  \\
         & + C_4\Bigg/\bigg(\exp \big(C_6 z^{C_7+\Omega_\text{in}}-C_8 \log (z) z^{C_7+\Omega_\text{in}}+C_{8} r_0 \log (z) \\
         & + C_{6} r_0\big) \cdot (C_{9}-\Omega_\text{in}) + C_5+\Omega_\text{in} \, \bigg) \, \Bigg)  \; \text{km} \, \text{s}^{-1} \text{Mpc}^{-1} .
    \end{aligned}
    \label{eq:sr_ht}
\end{equation}

Although this is a high complex expression, TuringBot was able to interpret this rate as the present unidimensional Hubble rate, scaled by a combination of factors that mainly depend on the redshift and the density, something analogous to the first Friedmann Equation found in FLRW, but here the inhomogeneity profile strongly contributes for the transverse hubble rate, as it can be seen in the denominator of the last term of this equation. The denominator contains linear combinations of $\Omega_\text{in}$, reflecting the analytic dependence of $H(r=0)$ on the central density in the constrained GBH model, while the exponential term seems to act like a kernel to create a smoother transition between the void profile (which could be written in terms hyperbolic functions, but that can be approximated, as SR seems to be doing here, by a series of exponential and logarithmic corrections in terms of the redshift). Once again, the values of the constants contrast with the cosmological parameter magnitude order attached to them.
   
\subsubsection{Radial Hubble rate}
An accurate approximate expression for the radial component of the Hubble rate is essential for the overall LTB framework, as it governs the light‑cone evolution and the relation between the radial and the transverse expansions. The Orchestrator found the following expression for $H_R$:
\begin{equation}
    \begin{aligned}
        \frac{H_R}{100} & = h \, \Bigg(
C_1 + (C_2 + z)^{C_3} - C_4\, \Delta_r \\
&\quad - \frac{C_5\,(C_6 - \Omega_\text{in})\, r_0}
{ C_7 \cdot \big(C_8 - C_9 r_0 +z\big)^{-1}
  - \tanh(C_{10}\,( r_0-\Delta_r ) - z) }
\Bigg) \\
&\text{km s}^{-1} \text{Mpc}^{-1}.
    \label{eq:sr_hr}
    \end{aligned}
\end{equation}

It is important to note that the radial component exhibits greater complexity due to its stronger dependence on the void profile, which encodes the inhomogeneous structure regime of LTB. Nevertheless, the expressions for $H_T$ and $H_R$ share several notable similarities. Although $H_R$ explicitly contains a hyperbolic tangent, its functional behavior is effectively exponential, as $\tanh(x) = (e^{2x}-1)/(e^{2x}+1)$. A comparable exponential dependence is also present in the lengthy denominator of the $H_T$ expression, indicating that both components are governed by similar underlying functional forms, albeit modulated differently by the inhomogeneity parameters. Moreover, the GBH general matter density $\Omega_M (r)$ (Eq.\ref{eq:GBH_omega_profile}) has hyperbolic terms as a function of the spatial cosmological parameters ($r_0$, $\Delta_r$). Our process was able to capture this behavior as $\tanh(f(r_0, \Delta_r))$, highlighting the SR capability of capturing physically motivated expressions.
Furthermore, the first three constants, $C_1$, $C_2$, and $C_3$, are of comparable magnitude in both expressions, with $C_2$ and $C_3$ differing only at the percent level. These constants consistently appear in the shared structural term $C_1+(C_2+z)^{C_3}$, suggesting that Symbolic Regression has identified a common baseline redshift dependence for both Hubble components. 

\subsubsection{Angular diameter distance}
To accurately model most cosmological observables, it is necessary to accurately compute the angular diameter distance $D_A(z)$. In LTB cosmologies with a central observer, this quantity is given exactly by Eq. \ref{eq:da_LTB} which shows that it depends on both the null geodesic solution and the transverse scale factor. Accordingly, whatever approximation is obtained by a method like SR, it must inherit the geometric complexity of the LTB spacetime, including its sensitivity to radial inhomogeneities. The best-derived formula for $D_A$ we obtained here was:

\begin{equation}
    \begin{aligned}
    D_A(z) & = D_H \Bigg[- C_9 (z + C_{10}) \; + \\
& C_1 \tanh\Bigg(
      \bigg(
        z \sqrt{
          (C_2 - \Omega_{\text{in}})
          (C_3 - C_4 r_0) + C_5
        } 
      \bigg)^{C_6 - C_7 \Omega_{\text{in}}}
    \Bigg) \Bigg]  \;  \text{Mpc.}
    \end{aligned}
    \label{eq:sr_da}
\end{equation}

It is noteworthy that the obtained expression is inversely proportional to the Hubble parameter, which aligns with units of distance, thus setting the characteristic distance scale of the approximated relation. Then, we can see that the linear term reproduces the expected low‑redshift behavior of $D_A$, while the hyperbolic tangent seems to capture the turnover and saturation of the angular diameter distance at intermediate redshifts, also capturing the dependencies on central density contrast $\Omega_{\text{in}}$ and and the void size $(r_0)$. Additionally, the components affecting $r_0$ exhibit a contrasting ordering relative to $r_0$, thereby normalizing this parameter, as we observe with the Hubble rates.

\subsection{Reconstruction accuracy}
Tab.~\ref{tab:tab_errors} summarizes the cumulative distribution of relative errors for each reconstructed function in our study. Columns 2 to 7 display the fraction of points reconstructed with relative errors less than $\pm0.1\%$, $\pm0.2\%$, $\pm0.5\%$, $\pm1.0\%$, $\pm2.0\%$, $\pm5\%$, and $\pm10\%$, respectively. The last two columns (Columns 8 and 9) show average and median relative errors for each case. 

Overall, these results indicate that the reconstruction accuracy improves rapidly as the relative-error threshold increases for all reconstructed functions. This is an indication of well-behaved error distributions without significant heavy tails. 
For most reconstructed quantities, more than $\sim 95\%$ of the points are recovered within $\pm 2\%$. Nearly all remaining points lie within the $\pm 5\%$ -- $\pm 10\%$ range. This provides evidence for a robust global performance of the symbolic regression approach.
The combination of high reconstruction fractions at loose thresholds and small average and median relative errors further suggests that large deviations are rare and do not dominate the error statistics.

The transverse scale factor $a_1$ exhibits a high precision, with over 99\% of predictions falling within a $\pm0.2\%$ error margin, and $100\%$ within $\pm0.5\%$. This indicates that the derived symbolic expression for $a_1$ closely matches the numerically computed values across the parameter space.

The radial scale factor $a_2$ and transverse Hubble rate $H_T$ also show high precision, with approximately 87\% and 74\% of predictions, respectively, lying within $\pm0.5\%$, and over 95\% within $\pm2.0\%$. In contrast, the radial Hubble rate $H_R$ and angular diameter distance $D_A$, which are more sensitive to radial inhomogeneity and redshift dependencies, achieve lower accuracy at strict thresholds with around $45\%$ and $55\%$ within $\pm0.5\%$, respectively. Nevertheless, both functions maintain robust agreement at broader thresholds, with $94\%$ of $H_R$ and over $98\%$ of $D_A$ estimates accurate to within $\pm10\%$.

These results highlight the capacity of Symbolic Regression to reproduce complex cosmological functions with high accuracy, particularly for quantities with simpler redshift and spatial dependence, while still yielding reasonable approximations for more intricate observables.

\section{Conclusions \label{sec:conclusions}}

Here, we presented an application of Symbolic Regression to obtain closed-form approximations for quantities in Lemaître-Tolman-Bondi cosmologies, focusing on the constrained Garcia-Bellido-Haugbølle parametrization. To do so, we used simulated data from the \texttt{bubble} code and a distributed SR workflow that we built around the TuringBot engine. We derived formulas for the radial and transverse scale factors, the associated Hubble rates, and the angular diameter distance. Our expressions reproduce the simulated data across the full parameter domain that was explored, and they resulted in mean relative errors less than 0.5\% for most quantities and 1.5\% for the radial Hubble rate.

These formulas obtained here reduce the need for costly integration of the LTB models, improving execution time by several orders of magnitude. This makes likelihood evaluations and model comparison using LTM models more tractable. Moreover, the closed-form nature of the SR approximations helps their interpretability and simplifies their incorporation into data analysis pipelines and/or inference methods. Our findings indicate that SR can produce reliable and interpretable surrogate models for inhomogeneous cosmologies, complementing the existing harder to interpret machine learning emulators and computationally costly traditional fitting approaches using data simulators

The method developed, including our parallel orchestration framework, can be directly applied and extended to other LTB parameterizations, to models with non-constant bang time, or to other categories of relativistic inhomogeneous cosmologies. We plan further studies to assess the performance of SR solutions in full Bayesian inference pipelines using real datasets and to explore the recovery of additional observables. The present study thus opens a path to generating analytic approximations for modeling relativistic inhomogeneities in the era of high-accuracy cosmology, as needed to make full use of modern datasets such as Euclid, Rubin/LSST, and the Nancy Grace Roman Space Telescope.

\begin{acknowledgements}
This work used the SDSC Expanse machine at San Diego Supercomputer Center through allocation PHY230141 (PIs Alberto Krone-Martins, António Silva, Celine Boehm) from the Advanced Cyberinfrastructure Coordination Ecosystem: Services \& Support (ACCESS) program, which is supported by U.S. National Science Foundation grants \#2138259, \#2138286, \#2138307, \#2137603, and \#2138296. We additionally thank the PHY230141 team, Hartley Tran, Esmerald Aliaj, Sang-Woo Jun, and Joshua Garcia for important discussions related to the development of this work. 

AC acknowledges support from the FCT research grant 2020.06644.BD and both FCT, Fulbright Portugal, and IIE for sponsoring a research visit of 6 months in University
of California-Irvine. The authors are also grateful to the Fundação para a Ciência e a Tecnologia (FCT) for the IA running research grants UIDB/04434/2020,
UIDP/04434/2020, EXPL/FIS-AST/1368/2021 (ML\_CLUSTER, DOI 10.54499/EXPL/FIS-AST/1368/2021) and PTDC/FIS-AST/0054/2021 (BEYLA, DOI 10.54499/PTDC/FIS-AST/0054/2021).

\end{acknowledgements}

%
%

\bibliography{references}
\bibliographystyle{aa}

\begin{appendix}
\section{The Orchestrator}
\label{appendix:code}

This appendix describes the orchestration framework developed to launch, manage, and coordinate multiple parallel TuringBot executions across multi-core computing environments. The framework was implemented in Python, evolving from an initial prototype developed in \texttt{R}, and was explicitly designed for deployment on high-performance computing (HPC) systems.

All large-scale experiments presented in this work were executed on the Expanse supercomputer, hosted at the San Diego Supercomputer Center (SDSC) and operated under the Advanced Cyberinfrastructure Coordination Ecosystem: Services \& Support (ACCESS) program. Expanse consists of 728 compute nodes equipped with dual AMD EPYC 7742 processors (64 cores per node), 52 GPU-enabled nodes with NVIDIA V100 GPUs, approximately 247~TB of system memory, and 824~TB of flash storage. Job scheduling and resource allocation are handled through the \texttt{SLURM} workload manager, with tasks submitted via the \texttt{sbatch} command using shell scripts written in \texttt{SLURM} syntax.

\subsection{Scalability Analysis}

To optimize the number of cores that each individual TuringBot process would be using, we first performed a tailored analysis of the scalability of TuringBot using our dataset and targeted parameter space on the Expanse system as a function of the number of concurrent threads. It is widely known that while increasing the number of threads enables additional parallel computations, excessive parallelism can introduce non-negligible overhead associated with thread management, memory contention, and context switching. Consequently in most real-life workloads, the performance does not increase linearly, and not even monotonically, with the number of threads.

\begin{figure}
    \centering
    \includegraphics[width=0.8\columnwidth]{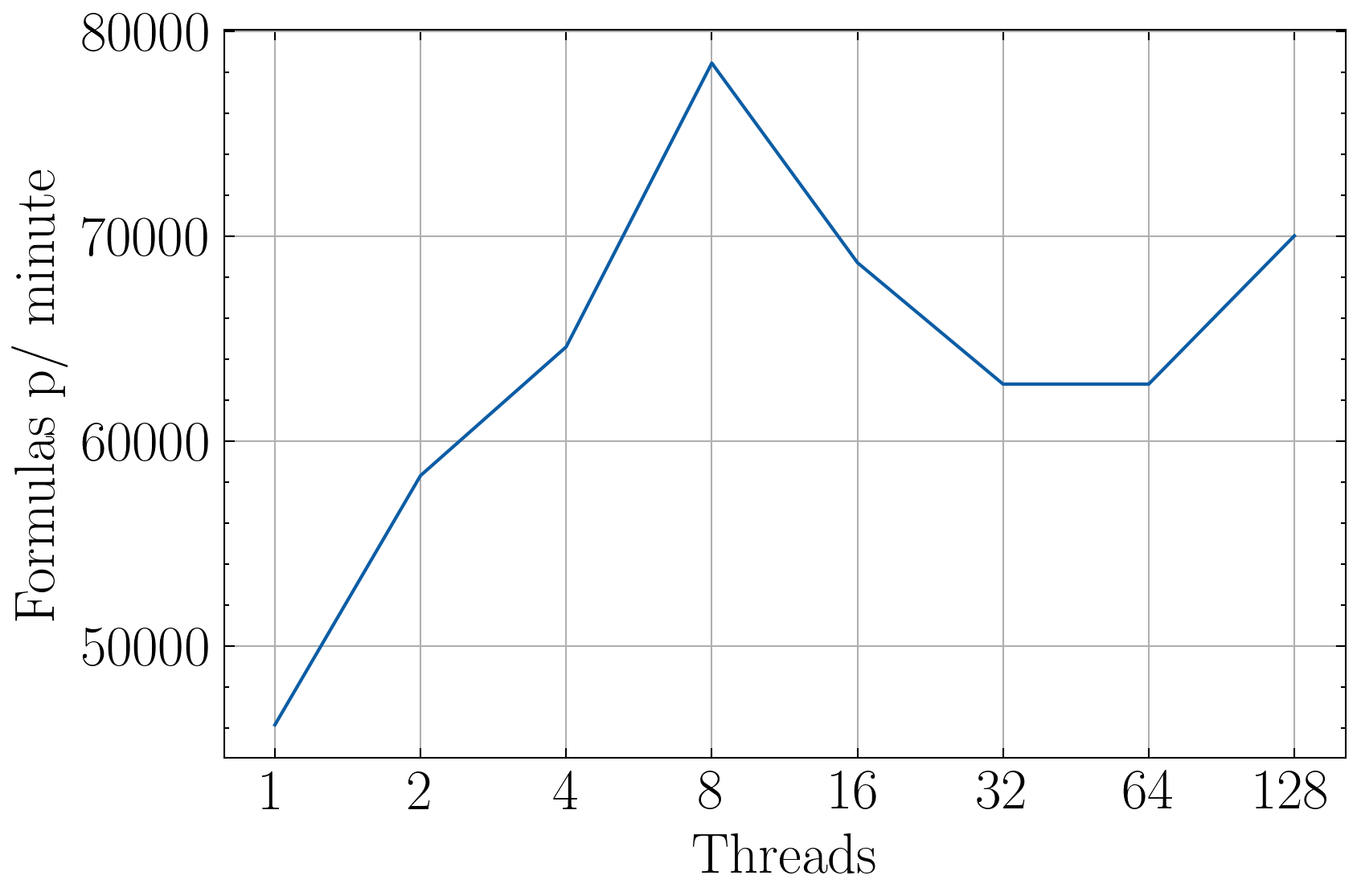}
    \caption{Scalability of TuringBot on the Expanse system. The figure shows the number of symbolic expressions generated per minute as a function of the number of threads.}
    \label{fig:thread_scale}
\end{figure}

Fig.~\ref{fig:thread_scale} reports the number of candidate symbolic expressions generated per minute as a function of the number of threads, for the datasets described in Sec.~\ref{sec:ltb_data}. The results indicate that performance saturates beyond a moderate number of threads. Based on these measurements, we selected eight threads as an optimal operating point, balancing computational throughput against resource utilization and scheduling overhead.

\subsection{Pipeline Architecture}
To efficiently exploit multi-core nodes and distributed cluster resources, we developed a dedicated orchestration software layer that manages multiple  TuringBot processes running in parallel. In cluster environments such as Expanse, TuringBot is executed within containerized environments, and here we use Singularity \citep{10.1371/journal.pone.0177459} and adopt container images in the Singularity Image Format (\texttt{.sif}).

The orchestrator framework we implemented adopts a client–host execution model. Computationally intensive tasks, including parallel TuringBot processes, are dispatched to compute nodes, while a dedicated control node is responsible for and handles workflow management, job coordination, and data aggregation. The Orchestrator framework is configured through a user-defined configuration file specifying resource allocation, execution parameters, container paths, and stopping criteria. One example of such configuration file can be found at Sect. \ref{ap:orch_file}.

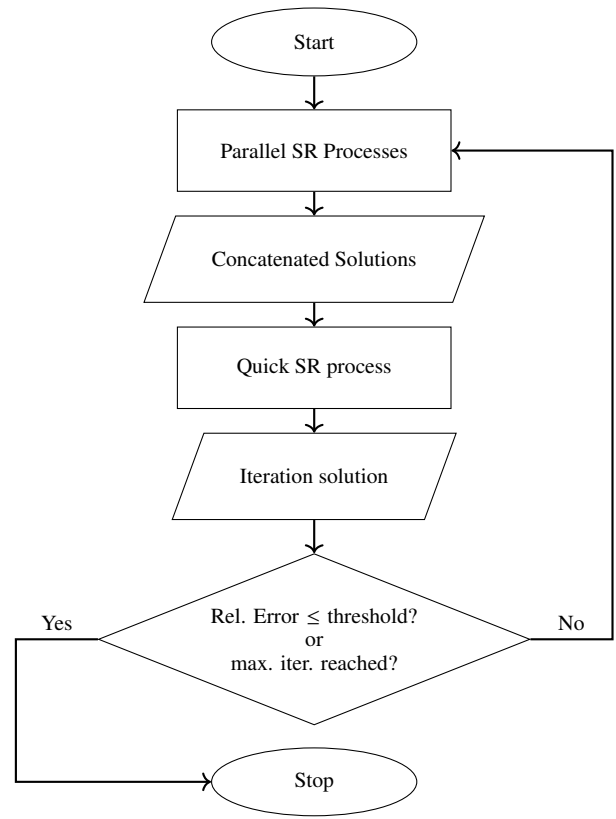
\begin{figure}
    \centering  
        \begin{tikzpicture}[scale=0.9, transform shape,
            node distance=1.6cm,
            every node/.style={font=\small},
            startstop/.style={ellipse, draw, minimum width=3cm, minimum height=1cm},
            process/.style={rectangle, draw, minimum width=4cm, minimum height=1.2cm},
            data/.style={trapezium, trapezium left angle=70, trapezium right angle=110,
                          draw, minimum width=4cm, minimum height=1.2cm, inner sep=15pt},
            decision/.style={diamond, draw, aspect=2.5, align=center,
    inner sep=-2pt},
            arrow/.style={->, thick}
        ]
        
        \node (start) [startstop] {Start};
        
        \node (parsr) [process, below of=start] {Parallel SR Processes};
        
        \node (concat) [data, below of=parsr] {Concatenated Solutions};
        
        \node (quicksr) [process, below of=concat] {Quick SR process};
        
        \node (itersol) [data, below of=quicksr] {Iteration solution};
        
        \node (decision) [decision, below of=itersol, yshift=-0.8cm, text width=4.5cm,
    align=center]
        {Rel. Error $\le$ threshold?\\ or \\ max.\ iter.\ reached?};
        
        \node (stop) [startstop, below of=decision, yshift=-0.5cm] {Stop};
        
        \draw [arrow] (start) -- (parsr);
        \draw [arrow] (parsr) -- (concat);
        \draw [arrow] (concat) -- (quicksr);
        \draw [arrow] (quicksr) -- (itersol);
        \draw [arrow] (itersol) -- (decision);
        
        \draw [arrow] (decision.west) -- ++(-1.2,0) node[midway, above] {Yes}
                       |- (stop.west);
        
        \draw [arrow] (decision.east) -- ++(1.2,0) node[midway, above] {No}
                       |- (parsr.east);
        
        \end{tikzpicture}
    \caption{Simplified flowchart illustrating the Orchestrator workflow used to automatically generate input files for the Symbolic Regression engine using multiple parallel batch process dispatches.}
    \label{fig:code_flowchart}
\end{figure}

A schematic overview of the Orchestrator pipeline can be seen in Fig.~\ref{fig:code_flowchart}. The internal workflow is organized into two main stages: 
\begin{enumerate}
    \item \textbf{Exploration stage.} Multiple parallel TuringBot instances are launched across the cluster and executed for a user-defined duration (Parallel SR Processes). Upon completion, the symbolic expressions discovered by each instance are aggregated into a single file by concatenating the individual outputs produced by each instance. This consolidated set of candidate equations serves as input for the subsequent stage (Concatenated Solutions).
    \item \textbf{Refinement stage.} A single TuringBot instance is executed for a shorter duration to refine and optimize the candidate expressions identified during the exploration stage (Quick SR Process). The resulting set of optimized expressions is then redistributed as input to each parallel process in the next iteration (Iteration Solution).
\end{enumerate}
These two stages are executed iteratively until either the maximum number of iterations specified by the user is reached or a predefined error threshold is satisfied. 

\subsection{Example of Orchestrator configuration parameters}
\label{ap:orch_file}
\begin{lstlisting}
[CPU]
nthreads = ; number of threads to be used in each of the longer processes
nprocesses = ; number of parallel processes
time = ; seconds for the longer process to be executed
ncicles = ; how many cicles
threshold = ; threshold value for the comparison made at the end of each cicle

[directory]
pwd = ; work directory (host)
folder_name = ; leave empty
folder_path = ${pwd}/${folder_name} ; do not edit
host_account = ; host account address
client_folder = ; output directory (client)

[sbatch]
job-name = ; slurm job name
output-name = ; name of slurm processess print output
memory = ; Expanse memory 
account = ; Expanse project
email = ; what e-mail slurm will send update messages to
shell_suffix =  ; full path and suffix of the batch files that the engine will create (host), ex: ${directory:folder_path}/shellfile
package_path = ; where the package is installed (host)

[TuringBot]
config_path = ; configuration file path of Turing Bot (host)
concat_suffix = ; full path and suffix of the concatenated files (host), ex: ${directory:folder_path}/Hralt_concat
equation_suffix = ; full path and suffix of the end of the cicle result file (host), ex: ${directory:folder_path}/Hralt_sol
training_path = ; full path of the train data (host)
test_path = ; full path of the test data (client)

use_prior_path = ; path of previous results to use (host). Leave empty if none
\end{lstlisting}

\end{appendix}

\end{document}